\documentclass[12pt]{iopart}

\usepackage{graphicx}

\usepackage{iopams}

\begin{document}

\title{Electronic properties of molecular solids: the peculiar case of solid Picene}

\author{Friedrich Roth$^{1}$, Matteo Gatti$^{2}$, Pierluigi Cudazzo$^{2}$, Mandy Grobosch$^{1}$, Benjamin Mahns$^{1}$, Bernd B\"uchner$^{1}$, Angel Rubio$^{2}$, Martin Knupfer$^{1}$}

\address{$^1$ IFW Dresden, P.O. Box 270116, D-01171 Dresden,
Germany}
\address{$^2$ Nano-Bio Spectroscopy group and ETSF Scientific Development Centre,   Dpto. F\'isica de Materiales, Universidad del Pa\'is Vasco,
  Centro de F\'isica de Materiales CSIC-UPV/EHU-MPC and DIPC,
  Av. Tolosa 72, E-20018 San Sebasti\'an, Spain}

\ead{f.roth@ifw-dresden.de}

\date{\today}


\begin{abstract}

Recently, a new organic superconductor, K-intercalated Picene with high transition temperatures $T_c$ (up to 18\,K) has been discovered. We have investigated the electronic properties of the undoped relative, solid picene, using a combination of experimental and theoretical methods. Our results provide detailed insight into the occuopied and unoccupied electronic states.

\end{abstract}

\section{Introduction}

Molecular crystals - built from $\pi$ conjugated molecules - have
been in the focus of research for a number of reasons. Within this
class of materials, almost every ground state can be realized at will,
spanning from insulators to semiconductors, metals,
superconductors or magnets. Due to their relatively open crystal
structure their electronic properties can be easily tuned by the
addition of electron acceptors and donors. In some cases, this
resulted in intriguing and unexpected physical properties. A
prominent example for the latter is the formation of metallic,
superconducting or insulating phases in the alkali metal doped
fullerides depending on their stoichiometry.\cite{gunnarssonbuch,rev4,forro01,knupfer01} In
particular the superconducting fullerides (e.g. K$_3$C$_{60}$)
have attracted a lot of attention, and rather high transition
temperatures above 30\,K could be realized. More recently,
interesting phenomena were observed in other alkali metal doped
organic molecular crystals as the observation of an
insulator-metal-insulator transition in 
alkali doped phthalocyanines.\cite{craciun06} However,
superconductivity with high transition temperatures similar to
those of the fullerides could not be observed in other molecular
crystals despite many research activities until recently, when superconductivity has been reported for another alkali metal doped molecular solid, K$_3$Picene, with transition temperatures up to 18\,K.\cite{mitsuhashi2010}

Here we present the first comprehensive investigation of the electronic properties of undoped solid picene using state-of-the-art experimental tools and first-principles many-body calculations.  Our results provide a detailed analysis of the occupied and empty states,
confirming unambiguously the presence of four flat quasi-degenerate conduction bands that could give rise to a high density of states around the Fermi level in the n-doped compound.
Moreover, the measured spectral properties can only be accounted for once the anisotropy of the  structure, 
local-field corrections and electronic correlations are considered.

\begin{figure}[h]
\centering
\includegraphics[width=0.5\textwidth]{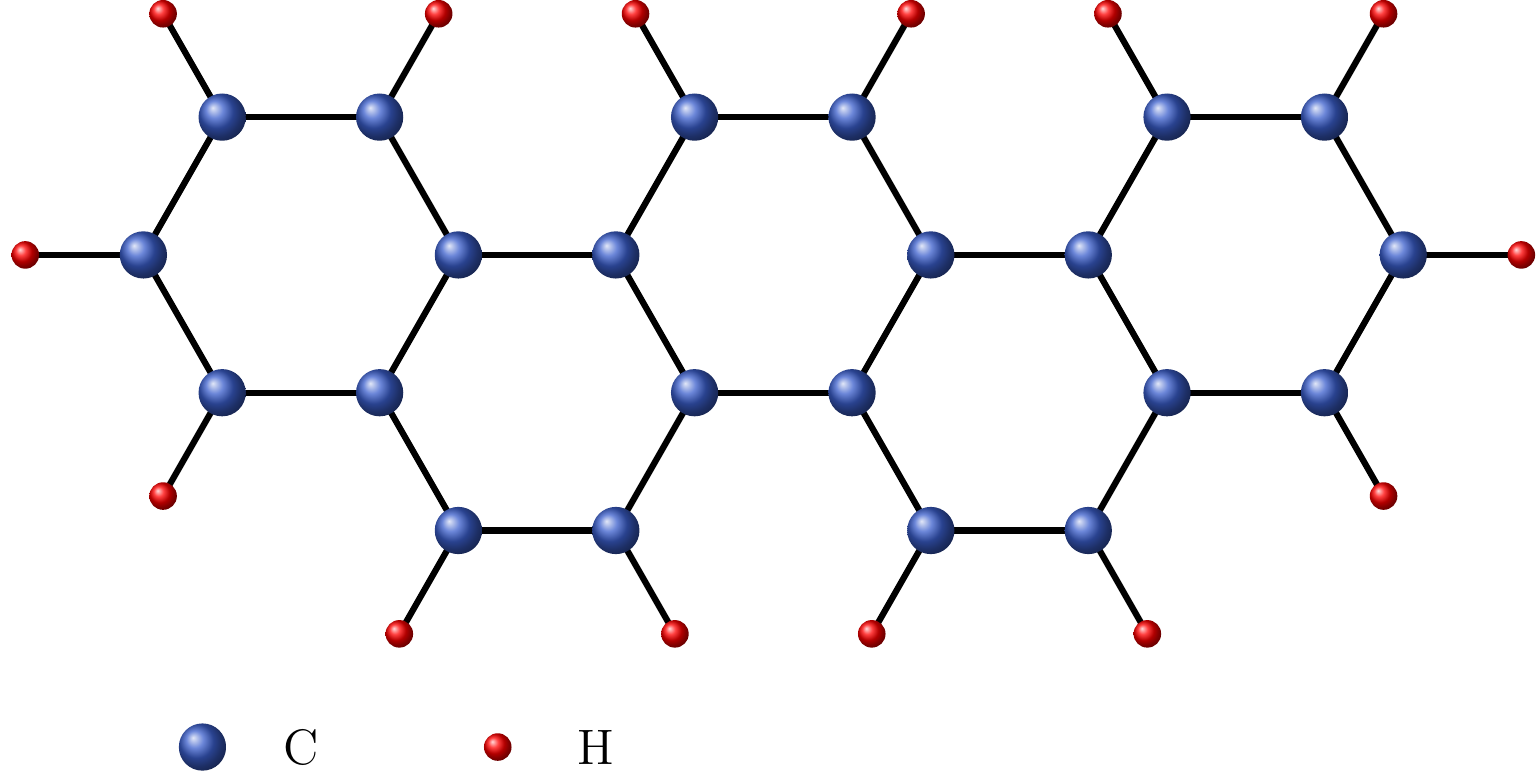}
\caption{Schematic representation of the molecular structure of Picene.} \label{f1}
\end{figure}

The Picene molecule is made by 5 benzene rings as depicted in Fig. \ref{f1}. In the condensed phase, Picene adopts a monoclinic crystal structure, with lattice constants $a$ = 8.48\,\AA, $b$ = 6.154\,\AA, $c$ = 13.515\,\AA, and $\beta$
= 90.46$^\circ$, the space group is P2$_1$, and the unit cell contains two inequivalent molecules.\cite{de1985} The molecules arrange in a herringbone manner which is typical for many aromatic molecular solids.

\par

\section{Methods}

For our photoemission studies, Picene films with a thickness of about 6 nm have been prepared by in-situ thermal evaporation with an evaporation rate of 0.1\,nm/min. onto a clean, polycrystalline gold substrate under ultra high vacuum conditions. The X-ray (XPS) and
ultra-violet (UPS) photoemission spectroscopy experiments have been carried out using a commercial PHI~5600 spectrometer, equipped with
two light sources. A monochromatized Al K$_{\alpha}$ source provides photons with an energy of 1486.6\,eV for XPS. Photons with an energy of
21.21\,eV (He discharge lamp) are used for valence band measurements. All UPS measurements
were done by applying a sample bias of -5\,V to obtain the correct, secondary electron cutoff which is required to determine the work function
and the ionization potential. The recorded spectra were corrected for the contributions of He satellite radiation. The total energy
resolution was about 350\,meV (XPS) and 100\,meV (UPS), respectively. The binding
energy (BE) scale was aligned by measuring the Fermi edge (0\,eV) and the Au4\textit{f}$_{7/2}$ emission feature (84.0\,eV) of the polycrystalline
gold substrate. For further details of the substrate cleaning and film characterization we refer the reader to previous
publications.\cite{peisert03}

For our investigations using electron energy-loss spectroscopy (EELS) in transmission we have prepared Picene films with a thickness of about
100\,nm by thermal evaporation under high vacuum onto single crystalline KBr substrates kept at room temperature. Thereby, two deposition rates, 0.2\,nm/min and 4\,nm/min, have been chosen to obtain films with different crystal orientations. Subsequently, the Picene films
were floated off in distilled water and mounted onto standard electron microscopy grids. Prior to the EELS measurements the films were characterized in-situ using electron diffraction. All observed
diffraction peaks were consistent with the crystal structure of Picene as given above. Moreover, the diffraction spectra revealed a pronounced texture, whereas the films grown with a deposition rate of 0.2\,nm/min showed a strong preference of crystallites with their $a,b$-plane parallel to the film surface, while those films grown with 4\,nm/min showed a substantial amount of crystallites with their $c$-axis in the film surface (see \cite{supmat} Fig. 6). The EELS measurements were carried out at room temperature using a 170\,keV spectrometer described elsewhere.\cite{fink1989} We note that at this high
primary energy only singlet excitations are possible. We have measured the loss function Im (-1/$\epsilon(\mathbf{q},\omega))$
($\epsilon(\mathbf{q},\omega)$ is the dielectric function) for a small momentum transfer $\mathbf{q}$ as well as the core excitation from the C1s core
level into unoccupied states. The energy and momentum resolution were chosen to be 85\,meV and 0.03\,\AA$^{-1}$, respectively. The measured loss
functions have been corrected for contributions from the elastic line.\cite{fink1989} For further details of the sample preparation procedure for
EELS and the experimental technique see \cite{fink1989,mkcpl}.

Experimental measurements have been complemented by first-principle electronic structure calculations.
Since density-functional theory (DFT) in local-density approximation (LDA) is known to underestimate band gaps  \cite{RMP},
quasiparticle energies have been calculated using the accurate $GW$ self-energy approximation \cite{hedin} of many-body perturbation theory,
where the self-energy is given by the product of the Green's function $G$ and the dynamically screened Coulomb interaction $W$.
The crystal structure has been optimized in LDA starting from experimental positions from Ref. \cite{de1985}
We have simulated the experimental loss functions in the random-phase approximation RPA \cite{RMP},
using norm-conserving pseudopotentials (with an energy cutoff of 40 Ha in the plane-wave basis set),
including 700 LDA bands in a 4x4x2  Monkhorst-Pack grid of $\mathbf{k}$ points.
Crystal local-field effects \cite{RMP} are taken into account by inverting a matrix $\epsilon^{-1}$ of rank 73 $\mathbf{G}$ vectors
in the reciprocal space.
For self-energy calculations we have used 7000 plane waves in the expansion of the wavefunctions, 350 empty bands,
a 6x6x4 Monkhorst-Pack $\mathbf{k}$-point grid, and a plasmon-pole model approximation in the calculation of $W$.
Quasiparticle energies are obtained as  first-order corrections to LDA eigenvalues. For comparison the calculated spectra have been convoluted with a Gaussian function of half width of 0.2\,eV for the PES and 0.075\,eV for the EELS.\\

\section{Results and Discussion}

\begin{figure}[t]
\centering
	\includegraphics[width=0.47\textwidth]{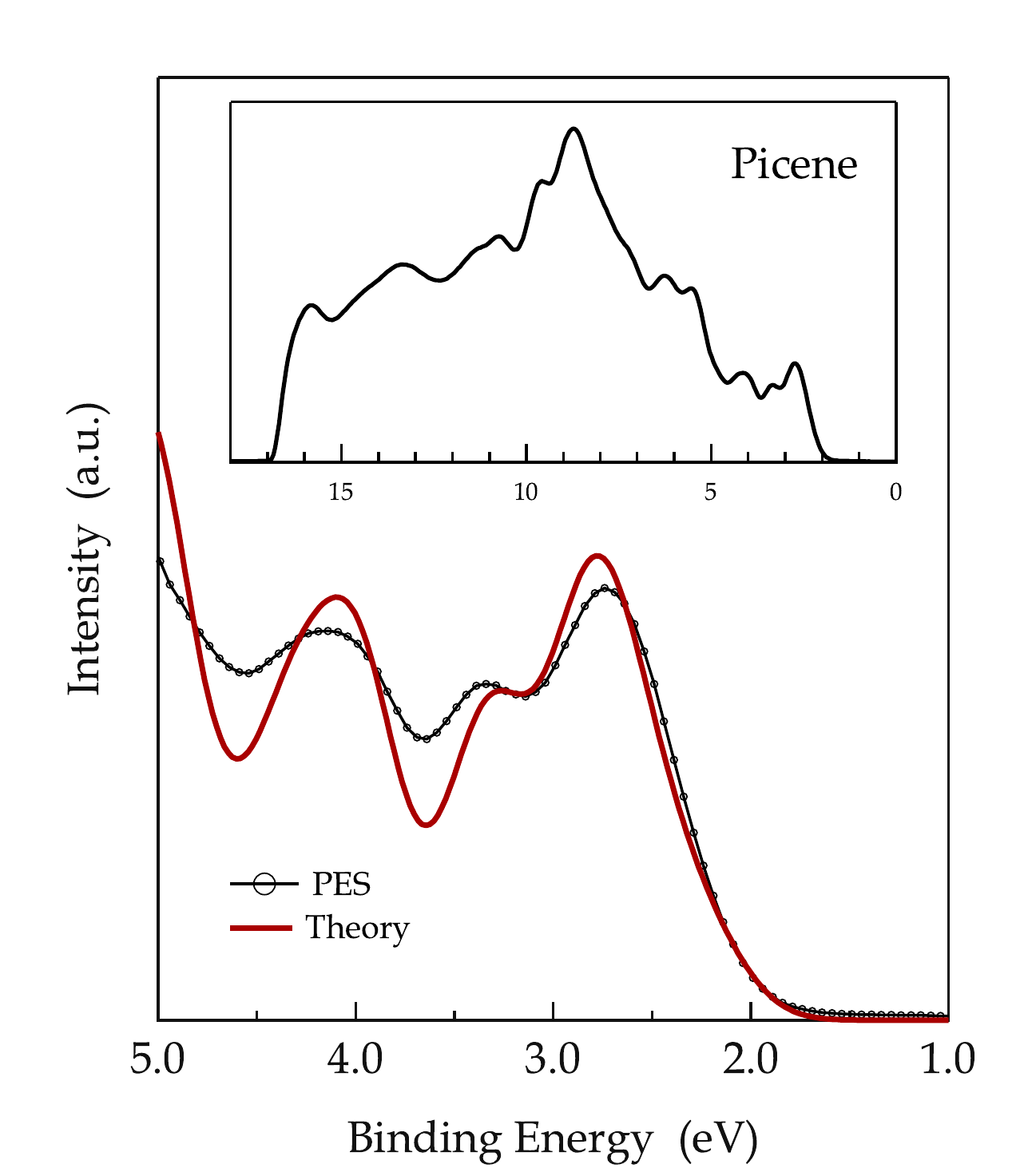}
	\includegraphics[width=0.47\textwidth]{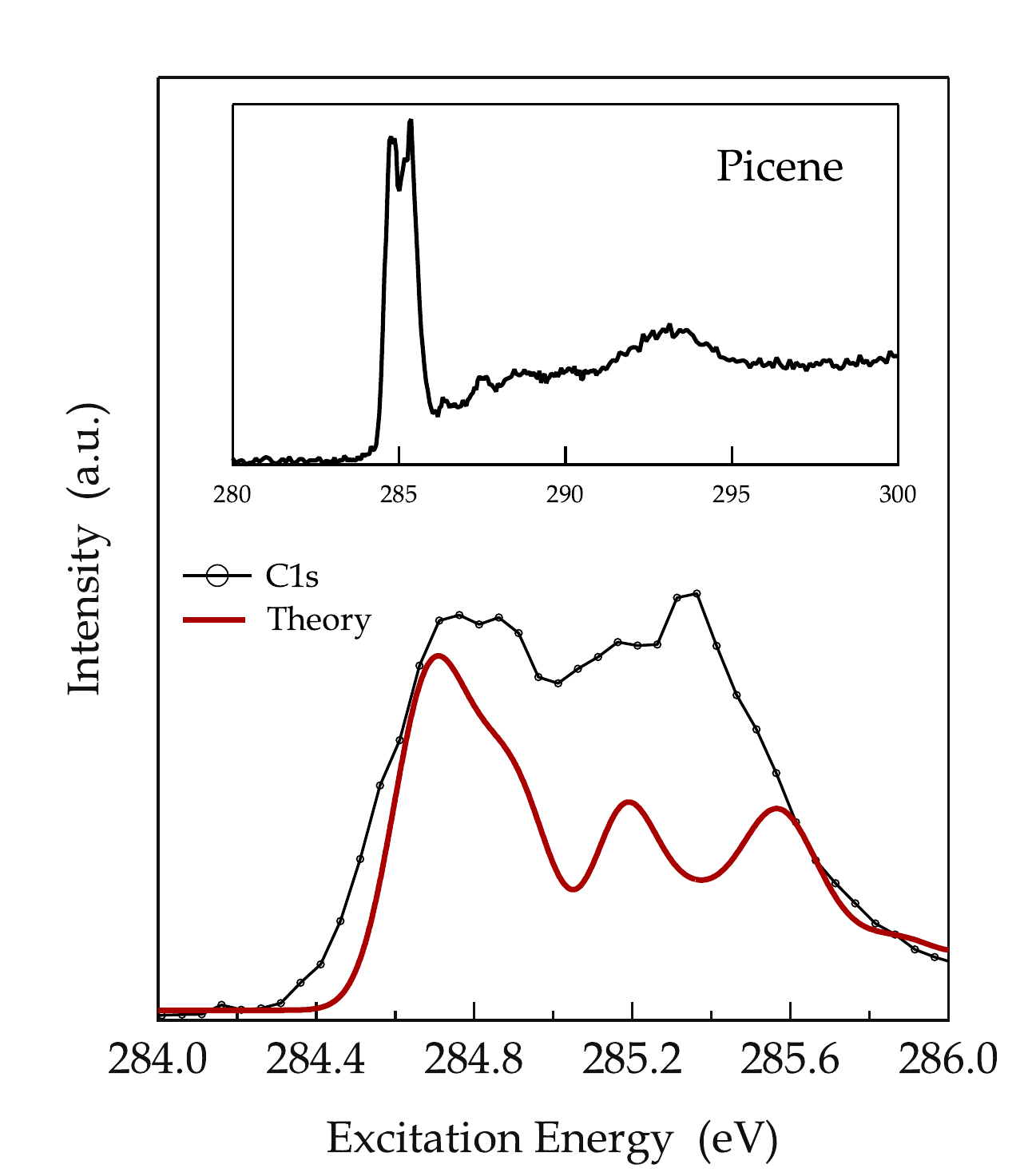}
\caption{Left panel: Valence band photoemission data of solid
Picene near the chemical potential. The inset shows the full
spectrum including the secondary electron cut-off. Right panel:
C1s excitation data of solid Picene measured using EELS. The inset
shows a much larger energy range. In the case of the C1s comparison the theoretical data have been shifted such that the first peaks coincide.} \label{f2}
\end{figure}

In the left panel of Fig. \ref{f2} we show the photoemission profiles of
Picene, which we compare with  theoretical densities of states (DOS), 
calculated within the accurate GW approximation of many-body perturbation theory.
The structures closest to the chemical potential (binding
energy = 0\,eV) arise from the $\pi$-derived highest occupied
molecular orbital (HOMO) of the Picene followed by deeper lying
electronic states (HOMO-1, HOMO-2 etc.). Upon solid formation, these orbitals form bands with a relatively small band width of about 0.5\,eV (see \cite{supmat} Fig. 1) 
since the interaction between the molecules in solid Picene is essentially van-der-Waals.
Our calculations  based on density-functional theory using the local density approximation (LDA) follow the results of Ref. \cite{kosugi2009},
but there is a major modification of the shape of the DOS once many-body correlation effects embedded in the $GW$ approximation are considered. Quasiparticle corrections change the positions
and the intensities  of the main peaks (see the detailed comparison in \cite{supmat} Fig. 2). 
The final result is in excellent agreement with the measured PES data.
Below about 6\,-\,7\,eV binding energy (BE) we find that the
$\sigma$-derived states additionally contribute to the
photoemission spectrum. The spectral sharpness of the
photoemission structures confirms that upon solid formation the
molecular electronic states of Picene remain relatively unchanged. 
Closest to the chemical potential, the photoemission data reveal
three maxima in the electronic density of states at 2.7\,eV, 3.35\,eV and 4.15\,eV, corresponding to the highest 8 valence bands  (see \cite{supmat}  Fig. 1). 

This observation demonstrates that the first occupied electronic
levels of Picene are quite close in energy.
The onset of the occupied electronic DOS is at about 2\,eV below
the chemical potential (or Fermi energy), which indicates a quite
large band gap of solid Picene. The ionization potential of solid
Picene as determined using the data in Fig. 2 is 6.4\,eV, i.e.
Picene is rather stable against oxidation. The work function of
Picene thus is 4.4\,eV.

In the right-hand panel of Fig. \ref{f2} the C1s core excitation spectra of Picene are depicted. Due to dipole selection rules, these data represent
transitions into empty C2p derived levels. In other words core level EELS is able to probe the projected unoccupied electronic DOS of carbon based materials.\cite{knupfer01,mkcarbon} We thus compare the experimental EELS with the unoccupied DOS calculated in the GW approximation.
Analogously to other $\pi$ electronic systems, the features below 291\,eV are caused by
excitations into $\pi^*$-derived electronic states. 
The step-like structure at about 291\,eV corresponds to the onset of transitions into
$\sigma^*$-derived unoccupied levels. The C1s core excitation spectrum of Picene shows a very sharp and dominating excitation feature right
after the excitation onset at 284.3\,eV, due to excitonic interactions with the core hole.\cite{mkcarbon,shirley,rehr} This excitation feature is characterized by a fine structure with maxima at 284.75\,eV, 284.85\,eV, 285.15\,eV and 285.35\,eV. These peaks are in very good correspondence with the theoretical data once $GW$ corrections are taken into account. From this analysis we can unambigously assign those structures to   
several unoccupied levels that are very close in energy (in the first 0.6\,eV we count 4 bands and 8 in the first 1.5\,eV - see \cite{supmat}  Fig. 1).
Compared to pentacene (also formed by five benzene rings, joined in a zigzag manner, instead than armchair-like as in picene),
these bands show an even smaller dispersion.\cite{tiago}
When doped with electrons, these states become occupied, and this quasi-degeneracy has been proposed to cause a
high DOS at the Fermi level in superconducting K$_3$Picene,
a situation that resembles that in fullerides and would be favorable for a relatively high transition temperature into the superconducting state. Moreover, this quasi-degeneracy of the conduction bands is also helpful to reduce the impact of electron correlation effects, i.e. to realize a metalic ground state similar to K$_3$C$_{60}$ \cite{gunnarssonbuch}, which is a necessary pre-requisite for superconductivity.

\begin{figure}[h]
\centering
\includegraphics[width=0.6\textwidth]{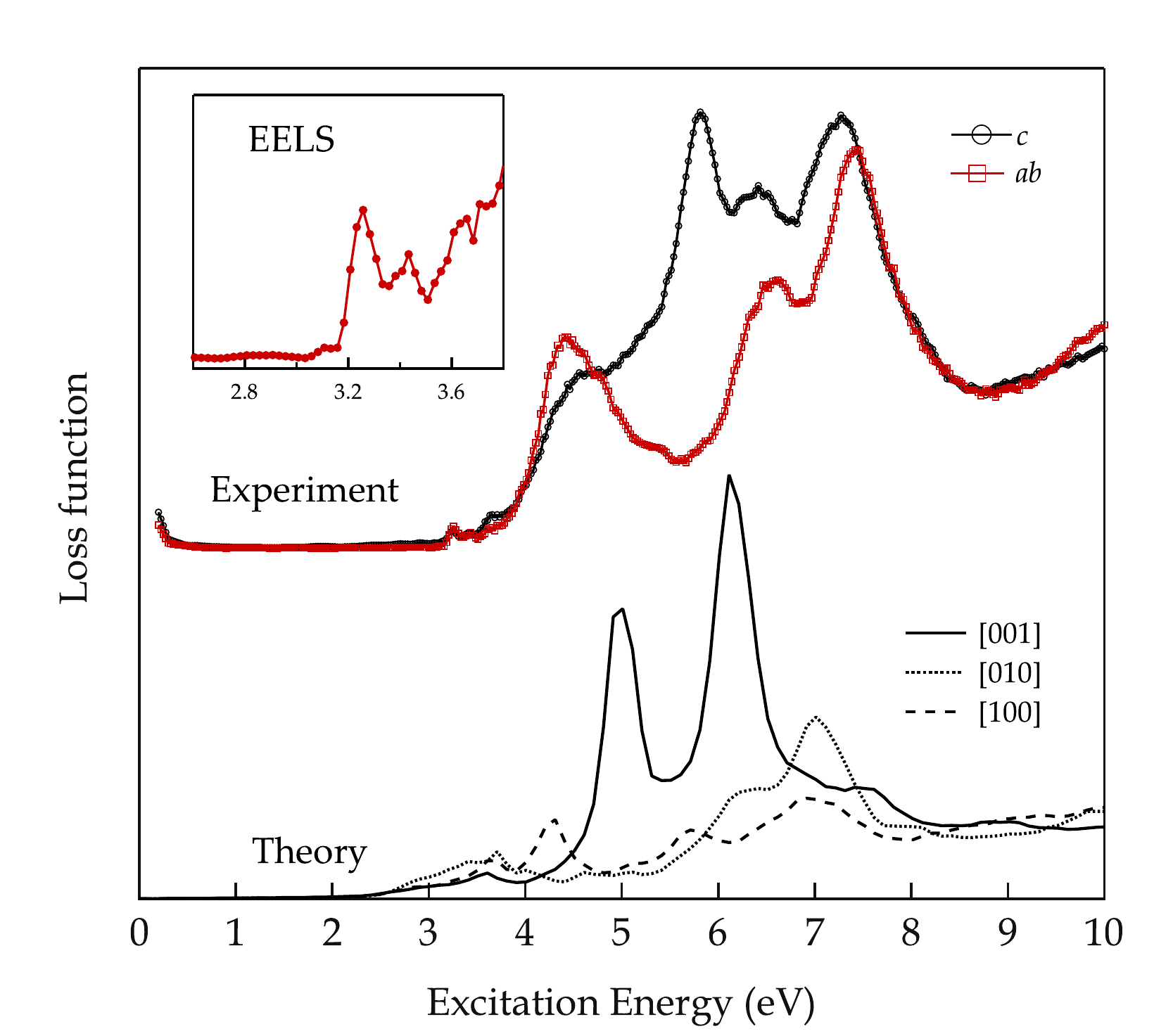}
\caption{Loss function of solid Picene at a small momentum
transfer of 0.1\,\AA$^{-1}$. The inset shows the spectrum in close
to the excitation onset. The experimental data represent excitations with predominant $a,b$ polarization (labelled $ab$, red curves) and with a strong contribution of excitations polarized along the $c$-axis (labelled $c$, black curve). } \label{f3}
\end{figure}

Finally, we present in Fig. \ref{f3} the loss function of solid Picene,
which provides insight into the electronic excitations of this
compound. The experimental data as presented in Fig. \ref{f3} are taken with a small
momentum transfer $\mathbf{q}$ of 0.1\,\AA$^{-1}$, which represents the
so-called optical limit.\cite{fink1989} Taking into account the
anisotropic molecular and crystal structure of Picene, it is
reasonable to expect an anisotropic loss function as well. 
The theoretical results for the loss function, calculated in RPA,
 match very  well the experimental measurements and
provide fundamental insight to interpret the spectra.

For a momentum transfer $\mathbf{q}$ parallel to the $z$ axis the loss function of Picene is dominated
by a broad structure at about 23\,eV, which is attributed to the volume
plasmon, a collective excitation of all valence electrons ($\pi+\sigma$ plasmon) (see \cite{supmat} Fig. 5).
The well-structured loss function below 10\,eV with clear maxima at
about 4.6\,eV, 5.8\,eV, 6.4\,eV, and 7.3\,eV in the experimental spectra is a signature of the
energetically sharp and well-defined molecular electronic levels
of Picene, which remain relatively unchanged going to the solid
state. 
The fact that the
observed maxima with high intensity are close in energy is also in
good agreement to the well structured data shown in Fig. \ref{f2} for the
electronic DOS, and we ascribe the excitation maxima in Fig. \ref{f3} to
excitations between the energetically close lying first occupied
and unoccupied electronic states of Picene.
With respect to the experimental spectra, the theoretical results show a slight underestimation of the peak positions
due to the fact that  the band gap opening due to GW corrections and excitonic effects - both absent in the present calculations -
do not  exactly compensate each other.
From the analysis of the theoretical spectra we find that in the spectrum with $\mathbf{q}$ parallel to the $z$ axis the first main structure  
at about 5\,eV is located in the continuum of single particle
excitations and is due to interband transitions (seen as peaks in the imaginary part of  $\epsilon$, see \cite{supmat}  Fig. 3).
Instead, the second peak at about 6.1\,eV occurs in correspondence to a zero of
the real part of $\epsilon$, where also the imaginary part is small. Therefore this structure can be assigned to a plasmon,
related to collective excitations with $\pi$ character.

Comparing the EELS spectra with momentum transfer along the three directions, we note that moving
from the $z$ axis towards the $x$ axis, the $\pi$ and $\pi+\sigma$ plasmons
undergo a blue shift and a red shift respectively. This can be understood in
terms of the one-particle excitations that originate them, as in layered systems like graphite and graphene.\cite{marinopoulos2002,marinopoulos2004} In fact, for $\mathbf{q}$ directed along the $z$ axis
(which is nearly parallel to the molecular main axis) $\pi-\sigma^*$ and $\sigma-\pi^*$ transitions
have very small oscillator strength. Therefore the main contributions in the low energy and high energy regions
of the spectra are given by $\pi-\pi^*$ and $\sigma-\sigma^*$ transitions
respectively. On the other hand, when $\mathbf{q}$  is directed along the other directions,
$\pi-\sigma^*$ and $\sigma-\pi^*$ transitions at intermediate frequencies increase considerably their oscillator strength.
As a consequence, the main structures of the loss function in the low-energy (the $\pi$ plasmon) and high-energy regions (the $\pi+\sigma$ plasmon)
undergo a blue and a red shift respectively.
The anisotropic character of the loss function is clearly visible also when we analyse the effect of crystal local-fields (LFE) on the spectra.
LFE are directly related to inhomogeneities in the induced charges (hence in the induced Hartree potential).
And, in fact, we find that LFE are almost negligible for $\mathbf{q}$ parallel to the $z$ axis, 
i.e. along the main molecular axis, where the charge distribution is quite homogeneous,
but they are stronger in the other directions as a manifestiation of the inherent anysotroy of  the molecular solid (see \cite{supmat} Fig. 4).

Zooming into the energy region around the excitation onset reveals an excitation onset in the experimental spectra,
i.e. an optical gap, of 3.15\,eV.
This onset also represents a lower limit for the band gap (or transport energy gap) of solid Picene.
While the fundamental band gap is severely underestimated by the 2.39\,eV LDA result  (see also Ref. \cite{kosugi2009}),
the GW band structure displays a direct quasiparticle
gap of 4.08\,eV (at the Z point of the Brillouin zone).  The GW result also implies that in Picene we should expect an exciton binding energy  larger  than
in the close relative pentacene, where it is less than 0.5\,eV \cite{tiago,hill2000,zahn06,kummer05},
and where also a smaller band gap (2.3\,eV) has been reported.\cite{tiago,kummer05,amy05}
The excitation onset of Picene is
followed by three rather weak electronic excitations at about 3.25\,eV, 3.4\,eV, and 3.65\,eV. Following previous optical studies of
Picene molecules in solution \cite{gallivan69}, we attribute these
low lying (singlet) excitations to those that are polarized
perpendicular to the long molecular axis, while at higher energies
excitations polarized along this axis contribut. Again, there is
a further clear difference to the electronic excitation spectrum
of pentacene, where the electronic excitation across the energy
gap has substantially more relative spectral
weight.\cite{schuster07}

\section{Summary}

In conclusion, we have studied the electronic properties of Picene using a combination of accurate experimental and theoretical spectroscopies.
Our results from photoemission and electron energy-loss spectroscopy have been analysed to clarify the effects of electronic correlation  and anisotropy in the dielectric response in order to understand its peculiar properties. With respect to Pentacene, Picene shows a larger quasiparticle gap, a larger exciton binding energy, and a higher density of unoccupied states close to the Fermi energy.

\ack We thank R. Sch\"onfelder, R. H\"ubel and S. Leger for technical
assistance, and we are grateful to the Deutsche
Forschungsgemeinschaft for financial support (KN393/5 and KN393/9).
This work has been  supported also by the Spanish MEC (FIS2007-65702-C02-01), ACI-Promociona (ACI2009-1036),
``Grupos Consolidados UPV/EHU del Gobierno Vasco" (IT-319-07), ETORTEK and
the European Union through  e-I3 ETSF (Contract: 211956) and THEMA  (Contract: 228539) projects. We acknowledge
support by the Barcelona Supercomputing Center, ``Red Espanola de Supercomputacion".
We have used Quantum Espresso \cite{pwscf}, Abinit \cite{abinit} and Yambo \cite{yambo}.

%%%%%%%%%%%%%%%%%%%%%%%%%%%%%%%%%%%%%%%
%%%%%%      SUPPLEMENTARY    %%%%%%%%%%
%%%%%%%%%%%%%%%%%%%%%%%%%%%%%%%%%%%%%%%

\clearpage
\setcounter{figure}{0}

\begin{center}
{\Large {\bf  Supplementary information}}
\end{center}

\vspace*{1.0cm}

\begin{figure}[ht]
\begin{center}
\includegraphics[width=0.49\columnwidth]{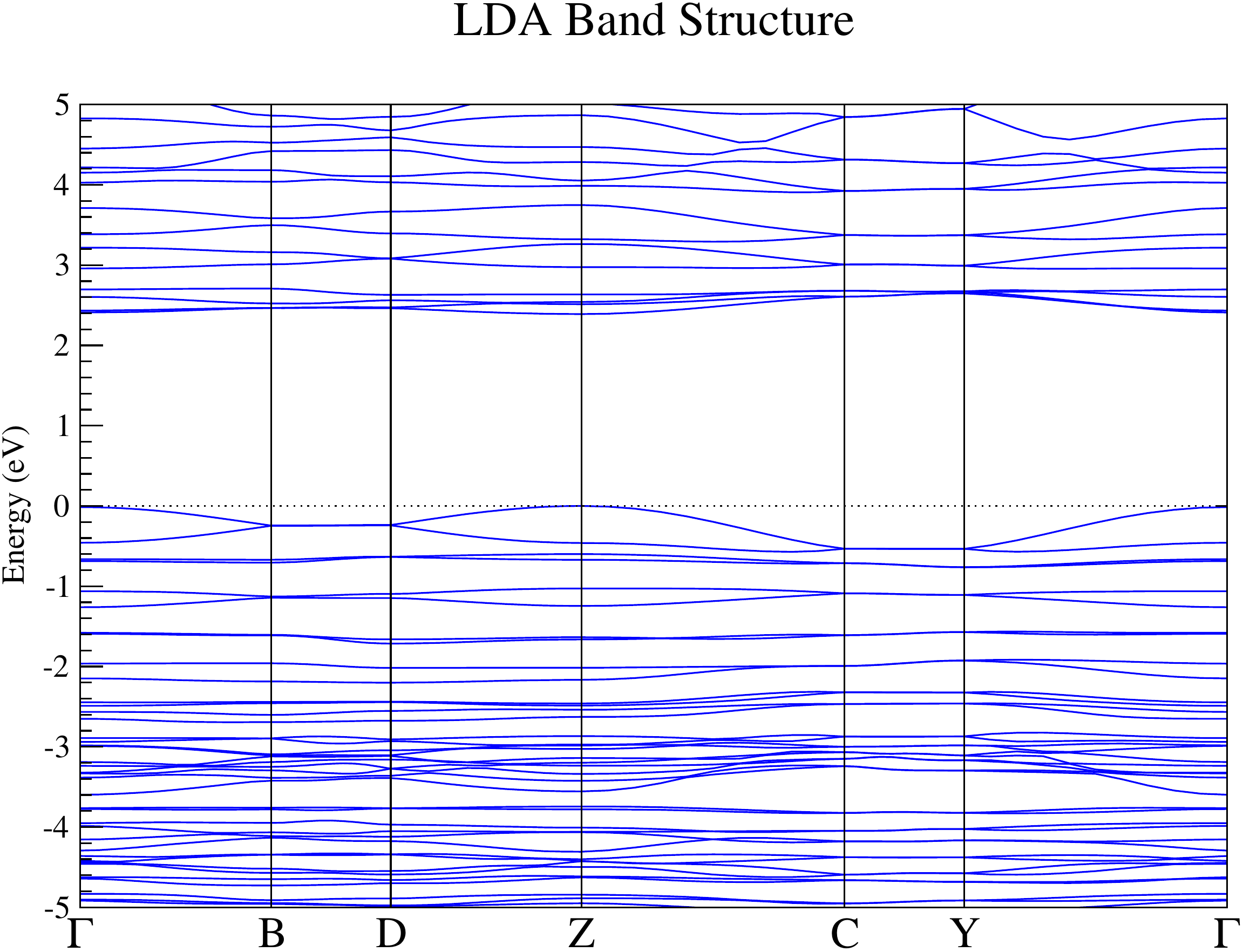}
\includegraphics[width=0.49\columnwidth]{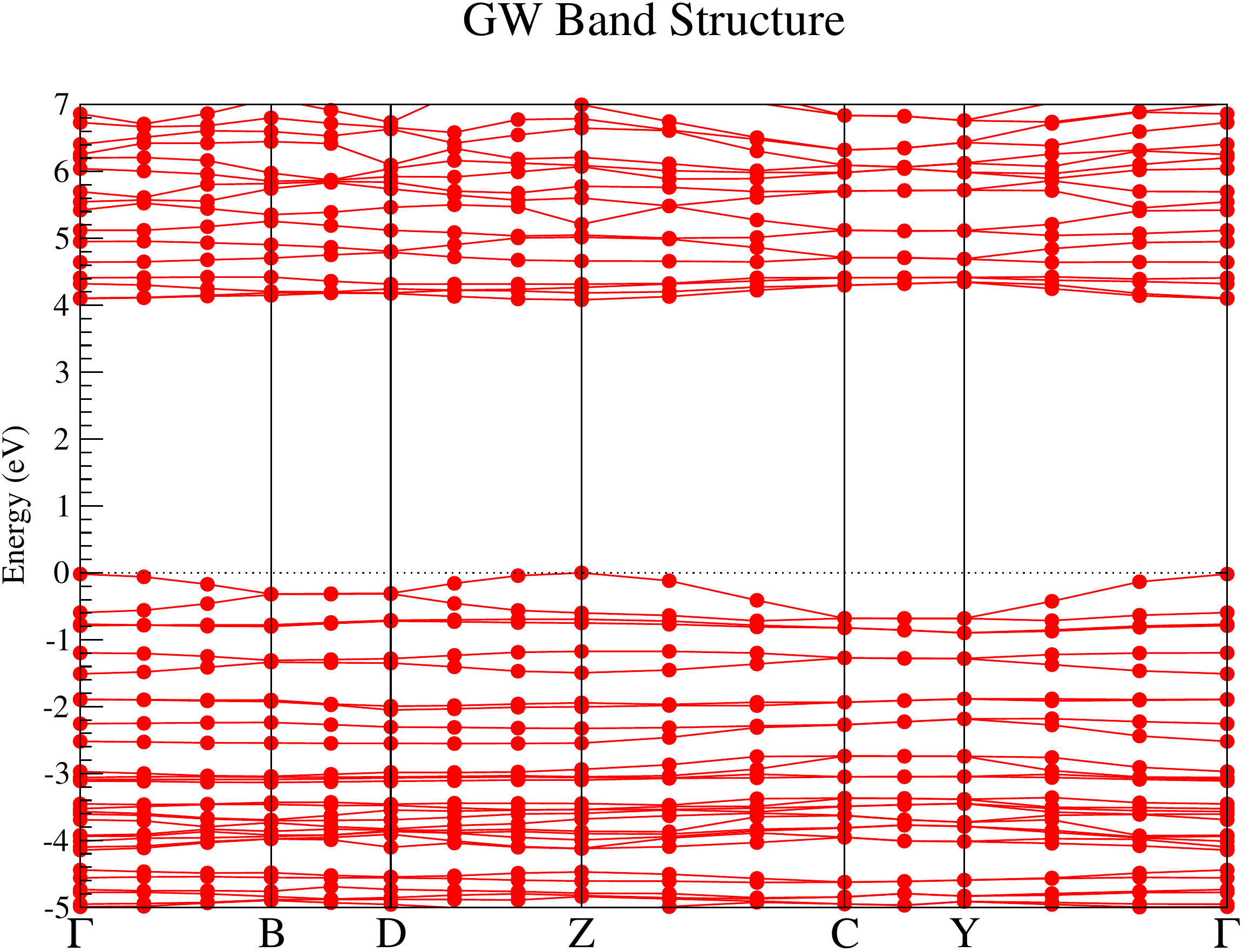}
\caption{Band structure of solid picene calculated in LDA (left panel) and in GW (right panel).
In the latter case, solid lines are a guide for the eye,
since GW corrections have been explicitly evaluated only  at $\mathbf{k}$ points corresponding to the dots.
The zero of the energy axis has been set to the corresponding top valence energy.}
\label{bands}
\end{center}
\end{figure}

\begin{figure}[ht]
\begin{center}
 \includegraphics[angle=270,width=0.66\columnwidth]{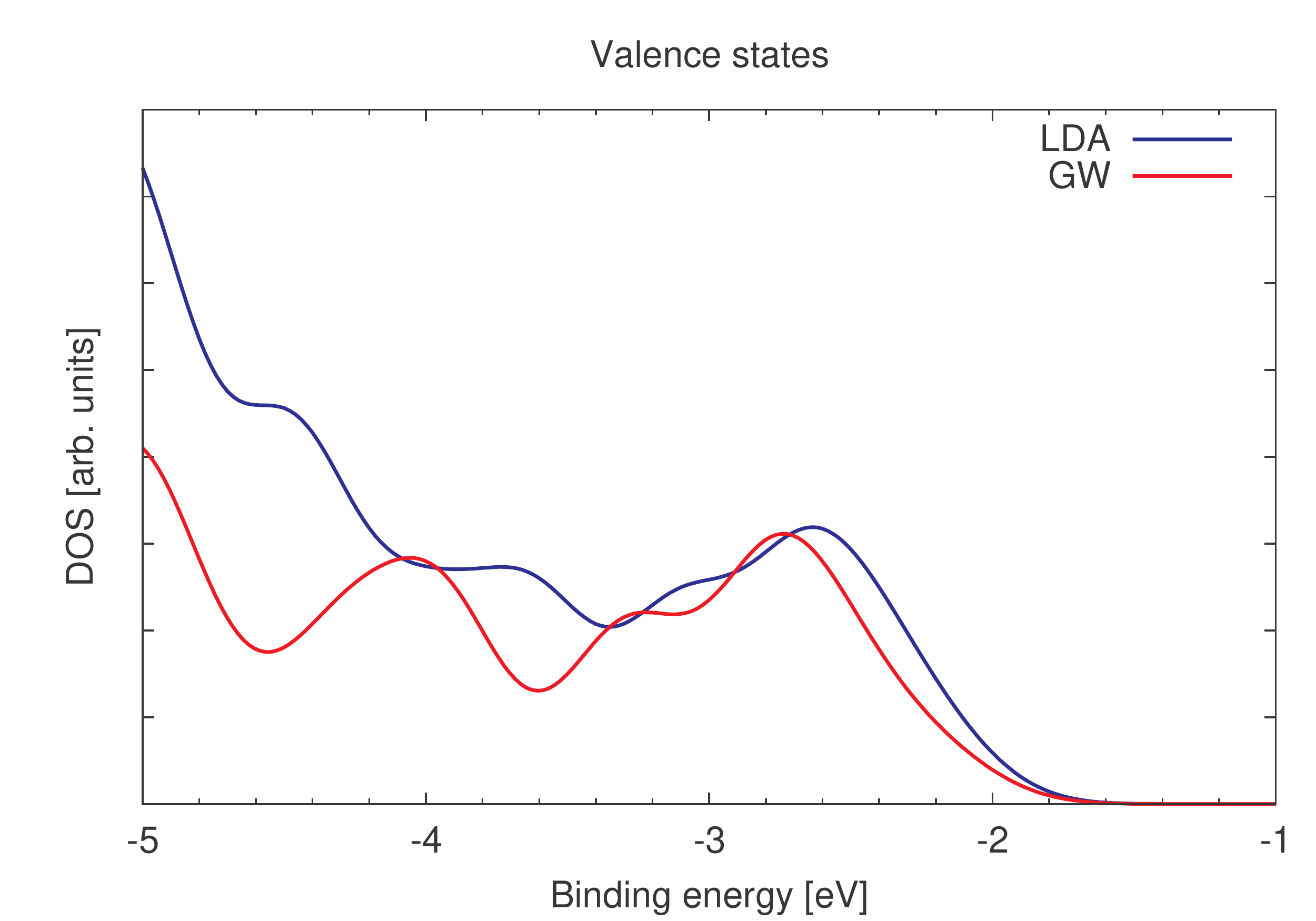}
\caption{Comparison between LDA and GW densities of valence states.}
\label{dos}
\end{center}
\end{figure}

\begin{figure}[ht]
\begin{center}
\includegraphics[width=0.38\columnwidth]{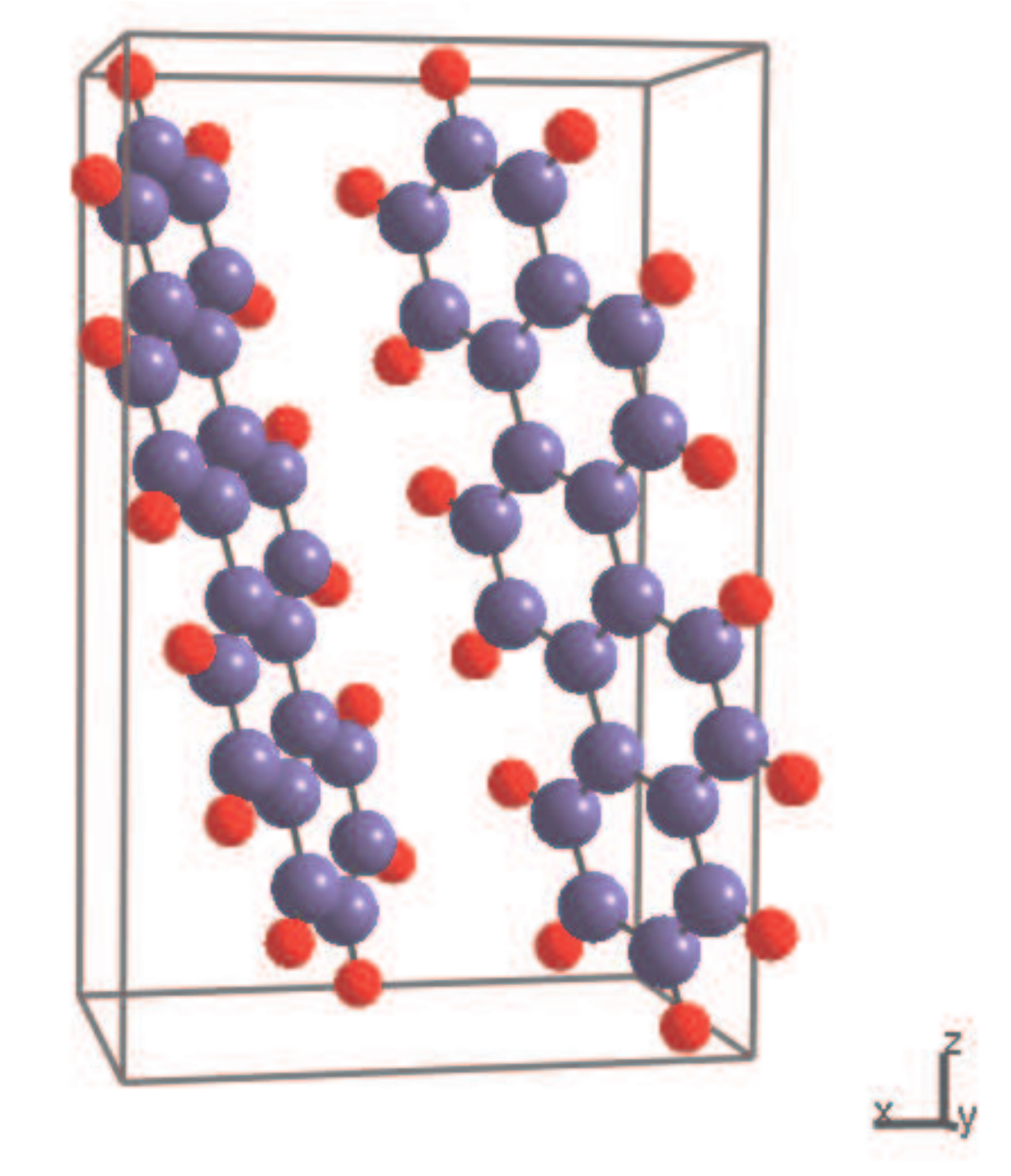}\quad
\includegraphics[width=0.55\columnwidth]{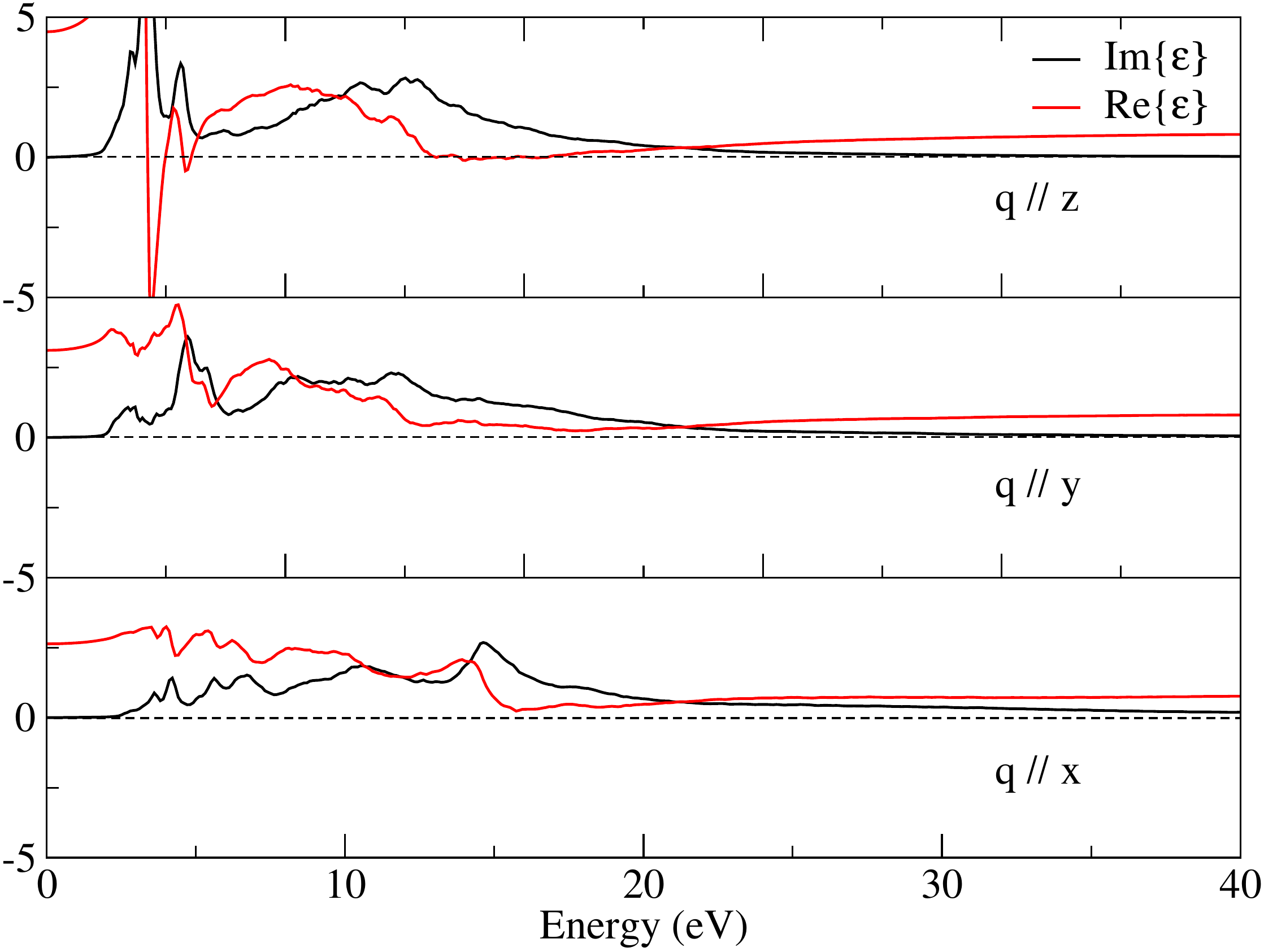}
\caption{(Left panel) Unit cell of solid picene (light blue: C atoms; red: H atoms). (Right panel) Real (red line) and imaginary (black line) part of the dielectric function of Picene evaluated along the three Cartesian directions}
\label{eps}
\end{center}
\end{figure}

\begin{figure}[ht]
\begin{center}
\includegraphics[width=0.66\columnwidth]{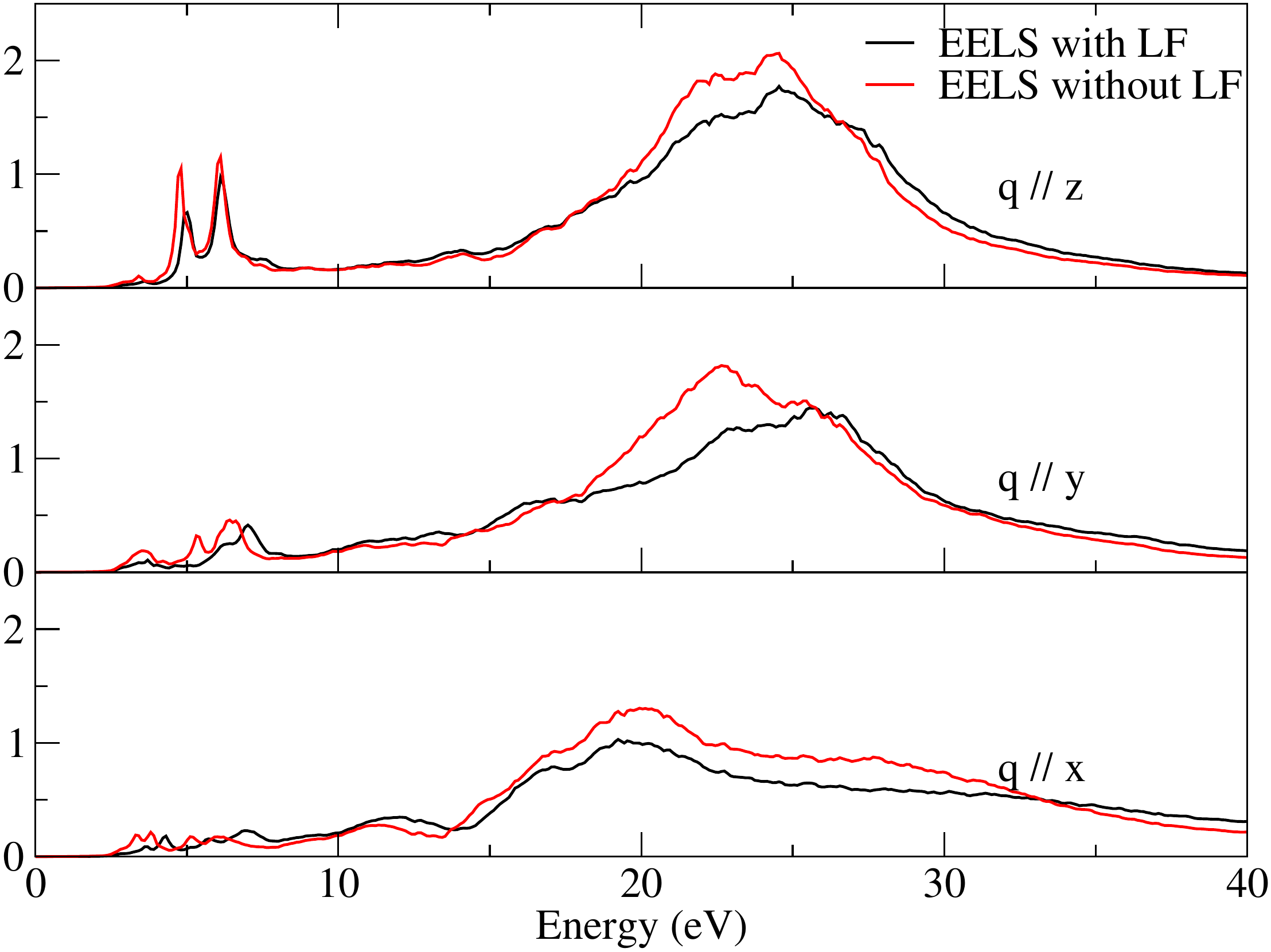}
\caption{Loss function of Picene with (black line) and without (red line) local fields evaluated along the three Cartesian directions}
\label{eels}
\end{center}
\end{figure}

\begin{figure}[ht]
\begin{center}
\includegraphics[width=0.65\columnwidth]{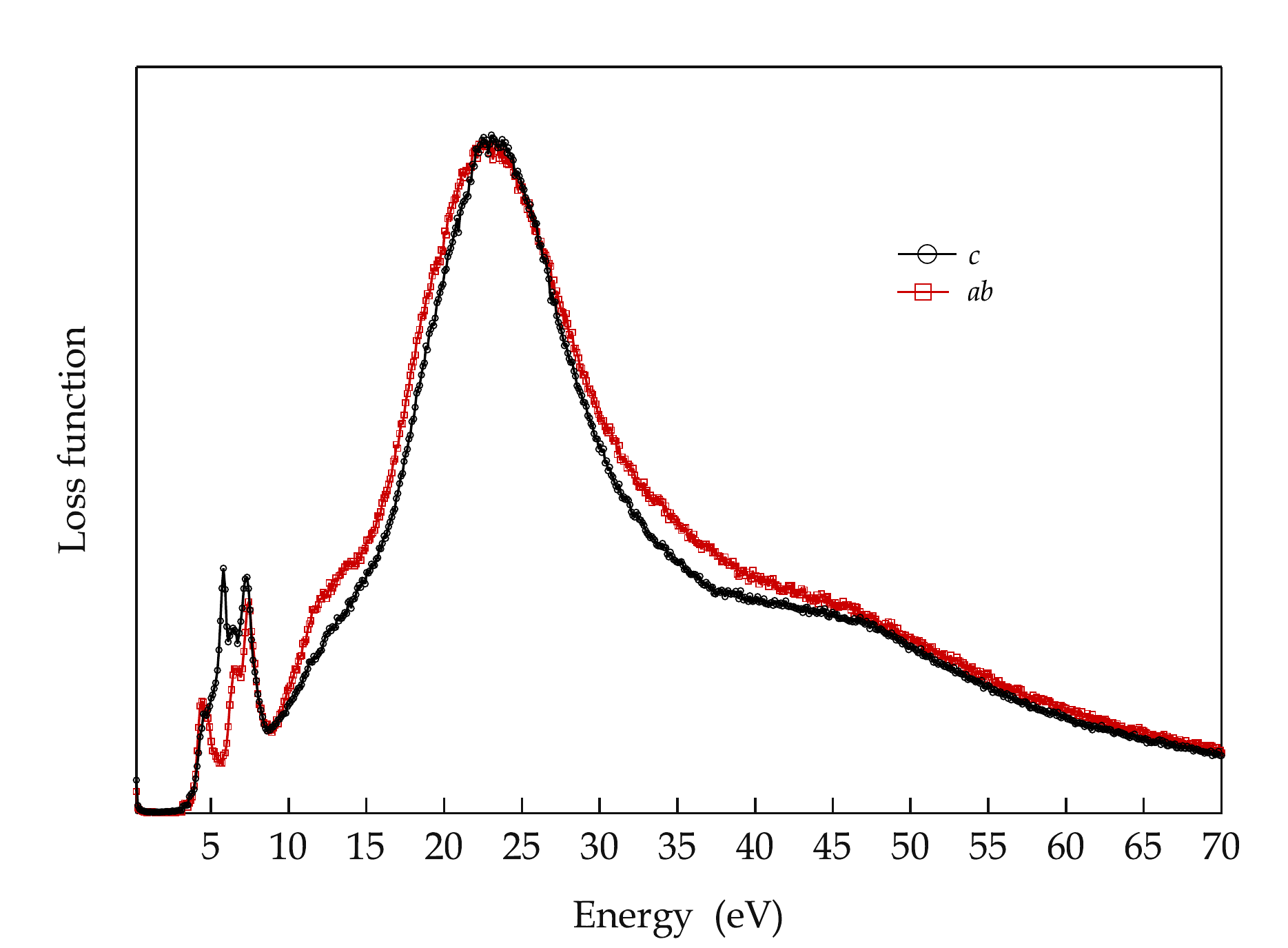}
\caption{Loss function of solid Picene at a small momentum
transfer of 0.1\,\AA$^{-1}$. The experimental data represent excitations with predominant $a,b$ polarization (labelled $ab$, red curves) and with a strong contribution of excitations polarized along the $c$-axis (labelled $c$, black curve). Note that the anisotropy which is seen in the calculations (Fig 3 above) for the $\pi+\sigma$ plasmon is hardly visible in the experimental data, which we attribute to the mixture of crystallite directions present in our samples. }
\label{loss_full}
\end{center}
\end{figure}

\begin{figure}[ht]
\begin{center}
\includegraphics[width=0.65\columnwidth]{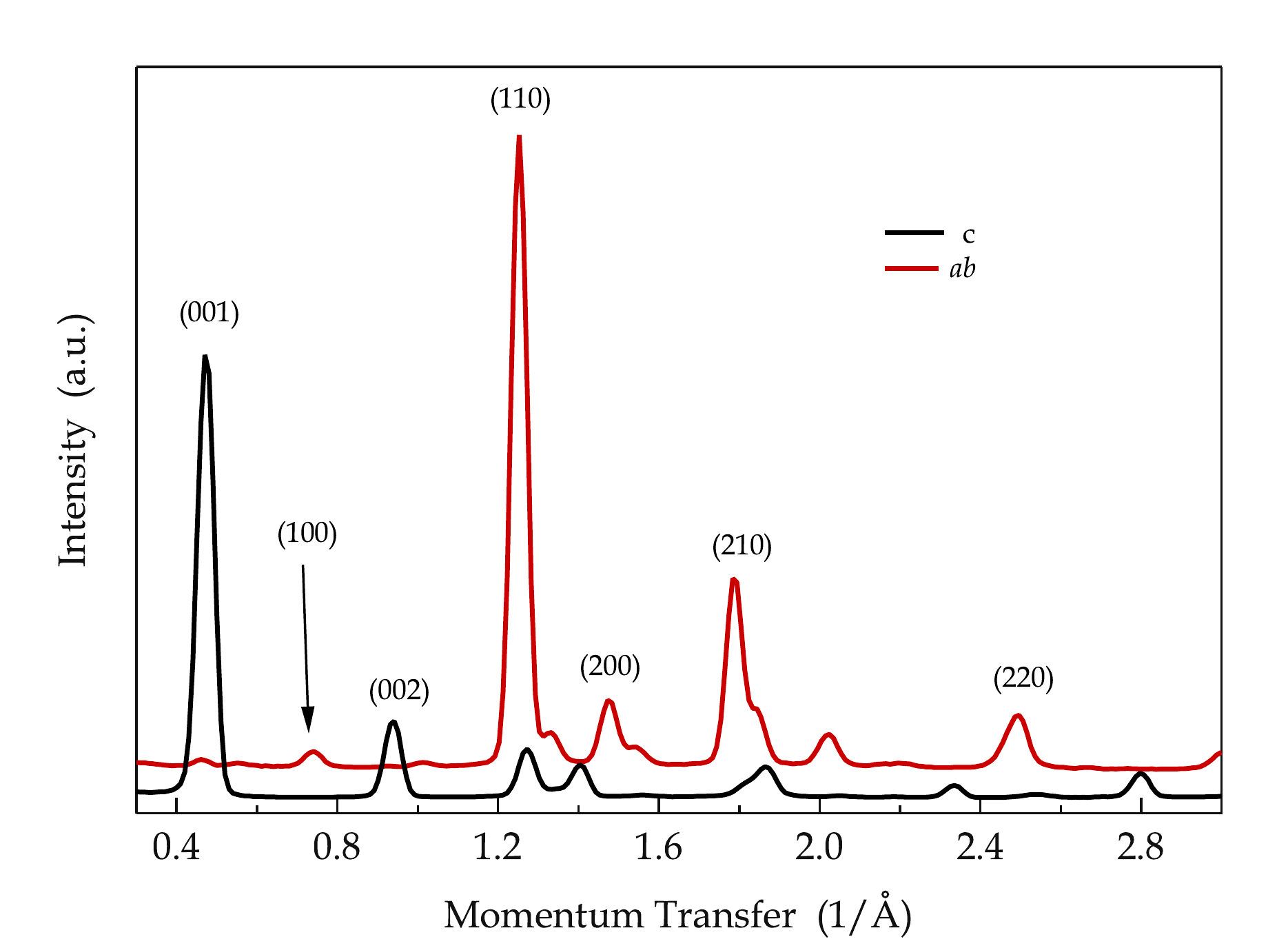}
\caption{Electron diffraction profils of solid Picene for the two different orientated films. The values in parentheses give the corresponding Miller indices and their positions are in fair agreement with the structural data described in. [A. De {\it et al.}, Acta Crystallogr. C {\bf 41}, 907 (1985)]}
\label{Bragg}
\end{center}
\end{figure}

\end{document}